\documentclass[doublecol,linenumbers]{epl2}

\newcommand{\vc}[1]{\mbox{\boldmath $#1$}}
\usepackage{amsmath}
\usepackage{amssymb}
\usepackage{graphicx}
\usepackage{dcolumn}
\usepackage{bm}
\usepackage{here}
\usepackage{color}
\usepackage{ulem}
\usepackage{MnSymbol}

\def\nn{\nonumber}
\def\bH{\begin{Huge}}
\def\eH{\end{Huge}}
\def\bL{\begin{Large}}
\def\eL{\end{Large}}
\def\bl{\begin{large}}
\def\el{\end{large}}
\def\beq{\begin{eqnarray}}
\def\eeq{\end{eqnarray}}

\def\eps{\epsilon}

\def\e{{\rm e}}

\def\>{\rangle}
\def\<{\langle}
\def\ll{\langle\!\langle}
\def\rr{\rangle\!\rangle}

\def \msd{{\Delta J^2}}

\title{Relation between irreversibility and entanglement in classically chaotic quantum kicked rotors }

\author{
Fumihiro Matsui{1}\thanks{E-mail: \email{fumihiro-matsui@xnea.net}} 
\and Hiroaki S. Yamada{2}\thanks{E-mail: \email{hyamada@uranus.dti.ne.jp}}
\and Kensuke S. Ikeda{3}\thanks{E-mail: \email{ahoo@ike-dyn.ritsumei.ac.jp}}
}
\shortauthor{F. Matsui, H. S. Yamada, K. S. Ikeda}

\institute{
\inst{1}{Department of Physics, College of Science and Engineering, 
Ritsumeikan University,Noji-higashi 1-1-1, Kusatsu 525-8577, Japan}\\
\inst{2}{Yamada Physics Research Laboratory, Aoyama 5-7-14-205, Niigata 950-2002, Japan}\\
\inst{3}{College of Science and Engineering, Ritsumeikan University, 
Noji-higashi 1-1-1, Kusatsu 525-8577, Japan}
}


\date{\today}
\abstract{
The relation between the degree of entanglement and time scale of 
time-irreversible behavior is investigated for classically chaotic quantum 
coupled kicked rotors by comparing the entanglement entropy (EE) and 
the lifetime of correspondence with classical decay of correlation, 
which was recently introduced. 
Both increase {\it on average} drastically with a strong correlation when 
the strength of coupling between the kicked rotors exceeds a certain threshold. 
The EE shows an anomalously large fluctuation resembling a critical fluctuation 
around the threshold value of coupling strength where the entanglement sharply 
increases toward full entanglement. 
In this regime it can be shown that, although the correlation is hidden, 
EE and the lifetime of {\it individual eigenfunctions} also have a positive 
correlation that can be seen via an another measure. 
}

\begin{document}
\maketitle


\def\H{\hat{H}}
\def\U{\hat{U}}
\def\Tr{\rm Tr}
\def\v{\hat{v}}
\def\J{\hat{J}}
\def\p{\hat{p}}
\def\q{\hat{q}}
\def\corr{C{\rm r}}

\def\vq{\hat{\vc{q}}}
\def\vp{\hat{\vc{p}}}
\def\vQ{\vc{Q}}
\def\vP{\vc{P}}

\def\T{{\cal T}}
\def\U{\hat{U}}
\def\tlife{\tau_L}
\def\avrtlife{\langle \! \langle \tau_L \rangle \! \rangle}
\def\dimN{N_{\rm dim}}

\def\Sm1{S_m^{(1)}}
\def\rhom1{\rho_m^{(1)}}

\section{Introduction}
Classically chaotic quantum systems can be considered as the simplest systems exhibiting 
apparently 
time-irreversible behavior \cite{prigogine}, and their quantum counterparts can also be the 
simplest systems realizing quantum irreversibility. 

Various examples of phenomena indicating evolution toward irreversibility, 
such as normal diffusion \cite{casati,fishman}, energy dissipation
\cite{ikedadissip}, energy spreading \cite{cohen} and so on have been presented. 
Moreover, by coupling with a proper classically chaotic quantum system as a 
``quantum noise source'', 
we can make a quantum system work as a quantum damper \cite{matsui}.  

However, quantum mechanical properties due to the quantum uncertainty principle 
prevent a quantum system from becoming irreversible in the same sense as 
its classical counterpart \cite{fishman,casati}.
Indeed unbounded chaotic diffusion of a classical system is
inhibited in its quantum counterpart. 
But if such systems are coupled even at 
classically 
negligible coupling strength, 
the diffusive motion exactly mimicking classical unbounded diffusion is recovered 
\cite{quantumclassical}.  Such diffusive motion 
exhibits characteristics of time-irreversibility identical to classical diffusive motion, 
that is complete decay of correlation and loss of past memory 
\cite{timerevexp,timerevexp-ballentine,benenti}.
A typical phenomenon showing the quantum recovery of chaotic irreversibility is the 
Anderson transition in the quantum standard map, which exhibits features of a critical 
phenomenon occurring due to cooperative effects \cite{casati2,delande}.
The drastic change of the coupled system is a reflection of the growth of 
entanglement 
among the constituent systems, which can be quantitively measured by the entanglement entropy
\cite{couplerotorentangle}.

We hypothesize that the spontaneous and cooperative recovery of classical chaotic 
irreversibility is a generic feature of a classically chaotic quantum system even if the 
system is bounded in a finite phase space region and so can not exhibit a truly diffusive
behavior. 
To quantitatively characterize the nature of time irreversibility 
grown up in quantum systems and their associated quantum states, we proposed 
a method to measure the {\it lifetime} of correspondence with the decay of 
temporal correlation as an indicator of the time scale on which a quantum 
system exhibits classical irreversible behavior \cite{matsuiletter}.
The aim of the present paper is to investigate the relationship between the entanglement
and the lifetime.

Irreversibility in quantum system has been explored directly by the time-reversal
experiment \cite{timerevexp,timerevexp-ballentine,benenti}, and further it has been 
extensively investigated by many authors in the context of fidelity \cite{fidelity1,fidelity2}. 
Our method shares some with the fidelity, but the advantage of our method is that it
enables us to observe the features of irreversibility related to the decay of correlation
over an extremely long time scale, and it enables us to measure the lifetime of 
classical irreversibility mentioned above.

\begin{figure}[htbp]
\begin{center}
\includegraphics[width=8.5cm]
{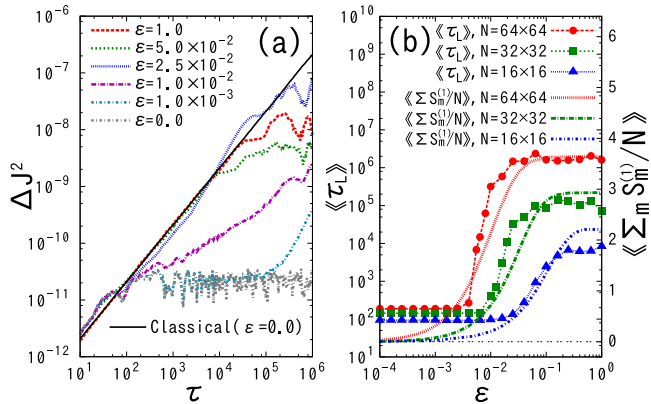}
\caption{
\label{fig1}(a) Some examples of time evolution of the mean square displacement(MSD) 
$\msd(\tau)$ at several $\eps$. The time scale
on which the quantum MSD can follow the classical MSD increases with $\eps$. 
Here $K=10$, $\eta=10^{-6}$ throughout this paper.
(b) The averaged lifetime $\langle \! \langle \tau \rangle \! \rangle$
 and the averaged mean entanglement entropy(EE) $\ll \sum_{m=1}^N \Sm1/N \rr$ 
as a function of $\eps$,　where $N_1=16,~32$, and $64$,
 where $\ll~~\rr$ means both $S$ and $S^T$ are ensemble-averaged.
The drastic enhancement of the former is strongly correlated with
the sudden increase of the latter. 
}
\end{center}
\end{figure}

\section{Model and method}
Our method is to convert the quantum motion in the bounded phase space of 
the object system to an extended motion in an infinitely extended homogeneous 
action space. If the motion of the system is classically chaotic and
correlation decays, the motion in the action space is a Brownian motion
in the classical limit \cite{matsuiletter}.

We showed that if the system's classical counterpart is fully chaotic 
with definitely decaying correlation, the lifetime of mimicking chaotic 
irreversible behavior can be measured as the time at which an evident deviation 
from the ideal Brownian-motion-like behavior takes place. 
The lifetime depends on the number of eigenstates composing the examined 
state, and the maximal lifetime is proportional to the square of Hilbert 
dimension of the system in the fully-developed entanglement regime \cite{matsuiletter}.

In the present paper we investigate the relation between the growth 
of irreversibility and the degree of entanglement in classically 
chaotic quantum systems. 
As a typical example, 
we take the coupled kicked rotors (CKR) system composed of two 
kicked rotors (KR) which are 
classically 
fully chaotic. The CKR is represented by
\begin{eqnarray}
&&  H(\vp,\vq,t)=(\p_1^2+\p_2^2)/2 \nn \\
&&     ~~~~+\delta_T(t)[V(\q_1)+V(\q_2)+\eps V_{12}(\q_1,\q_2)]
\end{eqnarray}
The two KR's interact via the interaction $V_{12}(\q_1,\q_2)=\cos(\q_1-\q_2)$ 
of strength $\eps$, where $\delta_T(t)=\sum_{n=-\infty}^{\infty}\delta(t-n T)$.
 We call the above system ``S''.
The KR's, say KR1 and KR2, are defined in the bounded phase space 
$(q_i,p_i) \in [0,2\pi] \times [0,2\pi]$ and so any diffusive motion does not happen.
The dimensions of the Hilbert space $N_1, ~N_2$ of the two KR's are equal $N_1=N_2$ 
and we take $\hbar=2\pi/N_i$. 
We take $T=1$ hereafter, although the notation ``$T$'' is remained in the mathematical
expressions.

Thus the dimension of 
the Hilbert space for the CKR is $N=N_1N_2=N_1^2$. The Arnold's cat map $V(\q)=-K\q^2/2$ 
and the bounded standard map $V(\q)=K\cos \q$ in the fully chaotic regime $K\gg 1$ are 
taken as examples of classically chaotic quantum systems.
The development of entanglement of eigenstates can be definitely controlled by the system
parameter.

Our method is to introduce a linear oscillator ``L'' which is very weakly coupled with 
the system S and converts S's quantum motion to a Brownian motion in 
the homogeneously extended action space of the linear oscillator.
L is represented by the angle-action canonical 
pair operators $\hat{\theta}$ and $\J=-i\hbar d/d \theta$ with the Hamiltonian $\omega \J$ of 
the frequency $\omega$ \cite{matsuiletter}.
\begin{eqnarray}
\label{hamil}
    \H_{tot}(t) = \H(\vp,\vq, t) + \eta \v(\vp)\hat{g}(\hat{\theta}) + \omega \J, 
\end{eqnarray}
where $\v(\vp)$ and $\hat{g}(\hat{\theta})$ are Hermitian operators, for which we take here
$\v(\vp)=\v(\hat{p_1})=\sin(\hat{p}_1)$ and $\hat{g}(\hat{\theta})=\cos(\hat{\theta})$.
$\J$ has the eigenvalue $J=j\hbar~(j\in {\bf Z})$ for the eigenstate 
$\<\theta|J\>\propto\e^{-iJ\theta/\hbar}$.
Observing the system at the integer 
multiples 
of the fundamental period $T(=1)$, that is $t=\tau T+0$ where
$\tau \in {\bf Z}$, then the one step evolution of CKR from $t=\tau T+0$
to $t+T$ occurs by the unitary operator 
$\U=\e^{-i[V(\q_1)+V(\q_2)+\eps V_{12}(\q_1,\q_2)]/\hbar}\e^{-i[\p_1^2+\p_2^2]T/2\hbar}$, 
and the Heisenberg equation of motion leads to the operator mapping rule, which immediately 
leads to the formula for the deviation of $\J$ from its initial value, namely, 
$\J(\tau)-\J(0)= \eta_\omega \sum_{s=0}^{\tau-1} \hat{f}_s$,
where $\eta_{\omega}=2\eta\sin(\omega T/2)/\omega$ and $\hat{f}_s=\v(\vp(s))\sin(\omega Ts+\hat{\theta}+\omega T/2)$.
Note that motion of $\vp(s) = \U^{-s} \vp \U^s$ is not influenced by the linear oscillator L in the limit of $\eta\to 0$, 
and further the expectation value $\<\J(\tau)\>$ vanishes if we take $|J=0\>$ as the initial state of L. 
This formula tells that, in the classical limit, $\J$ exhibits a Brownian motion driven by the chaotic force $\hat{f}_s$.
The mean square displacement  (MSD) $\msd = \<(\J(\tau)-\<J(\tau)\>)^2\>$ is then 
\begin{eqnarray}
\label{msd}
\msd(\tau) = \sum_{s=0}^{\tau-1} D_\omega(s),~~ D_\omega(\tau)=D\sum_{s=-\tau}^\tau \corr_\tau(s)\cos(\omega Ts)
\end{eqnarray}
where  $\<...\>$ means the expectation value with respect to the initial wavepacket 
$|\Psi_0\> \otimes |J=0\>$, 
where $|\Psi_0\>$ is the initial state of the system S, 
and $D =2\eta^2\sin^2(\omega T/2)/\omega^2$.  
$\corr_\tau(s)=(\<\Psi_0|\v_{\tau} \v_{\tau-s}|\Psi_0\> + c.c.)/2~~(s\geq 0)$
and $\corr_{\tau}(-s)=\corr_{\tau}(s)$, where $\hat{v}_{\tau}=\U^{-\tau}\hat{v}\U^\tau$, 
is the autocorrelation function.

Since the classical CKR system is an ideally chaotic system, it exhibits persistent diffusion, 
the autocorrelation function decays exponentially with the Markovian property, 
and $D_{\omega}(\tau)$ converges to a constant value $D_{\omega}^{(cl)}$ in the limit 
$\tau \rightarrow \infty$, which are manifestations of continuous loss of memory 
in classical chaos. However, in quantum systems, convergence of $D_\omega(\tau)$ to 
a finite value occurs only for a finite $\tau$. Indeed the quantum $D_\omega(\tau)$ is 
explicitly represented by
\begin{eqnarray}
 \label{diffcon} 
  D_\omega(\tau) &\sim& D \sum_{s=-\tau}^{\tau}\sum_{m=1}^M|C_m|^2 \sum_{n=1}^N |\<m|\v|n\>|^2 \nn \\
       &&  \times \cos((\gamma_m-\gamma_n)s)\cos(\omega Ts)
\end{eqnarray}
using quasi-eigenstate $|m\>$ and eigenangle $\gamma_m$ of the evolution operator $\U$ of S
i.e., $\U|m\>=e^{-i\gamma_m}|m\>$, and the initial state $|\Psi_0\>=\sum_{m=1}^{M}C_m|m\>$, 
\footnote{Eq.(\ref{diffcon}) takes a rather simple form, because we suppose the 
``diagonal approximation'' neglecting the terms of $C_{m'}^*C_{m}$ with $m' \neq m$.
It is confirmed as an extremely good approximation for {\it chaotic} KR's.}
Thus the quantum $D_{\omega}(\tau)$ 
is a sum of periodic oscillations and therefore vanishes if it is averaged 
on a long enough time scale, which means that the stationary diffusion is 
suppressed beyond a certain time scale $\tlife$. 
We regard this time scale to be the lifetime of correspondence with classical diffusion, and, equivalently, 
{\it the lifetime of correspondence with the classical decay of correlation}. 
The lifetime $\tlife$ indicates the maximal time scale on which the quantum time-irreversibility is observable.

Equations (\ref{msd}) and (\ref{diffcon}) are not convenient for numerical computation 
over an extremely long time scale. 
Practically, with the method of wavepacket propagation starting from the initial state 
, we compute $D_\omega(\tau)$ from the data of MSD \cite{matsuiletter}.
With these data we decide $\tlife$ as the time at which
there occurs a significant deviation from the linear growth of MSD, which is the most definite feature of chaotic irreversibility. \\
(i) First we decide the diffusion exponent $\alpha(\tau)$ such 
that $\msd(s) \propto s^{\alpha(\tau)}$ 
defined for $s$ in an appropriate 
interval around $\tau$ with the same width in logarithmic scale.\\
(ii) Next, we decide $\tlife$ as the 
time at which $\alpha$ deviates from $1$, specifically the first time step 
that $|\alpha(\tlife)-1|>r$, where we choose $r=0.5$ in practice.\\
(iii) The value of $\tlife$ is very sensitive to the choice of initial condition and also 
to the parameter $\eps$ in the regime of transition to full entanglement. \\
According to the procedure of \cite{matsuiletter}, we
can eliminate fluctuations of the lifetime due to such sensitivity: 
we add classically negligible small perturbation such as $\xi_{iR}\cos(\q_i-q_{iR})$ 
of order $O(\hbar)$, 
to the potential $V(\q_i)$ 
and take the average of $\tlife$ over
the ensemble of the potential parameter $q_{iR}~(i=1,2)$. We refer 
to the result of this procedure 
hereafter as the average lifetime $\avrtlife$. \\

\begin{figure}[htbp]
\begin{center}
\includegraphics[width=8.0cm]
{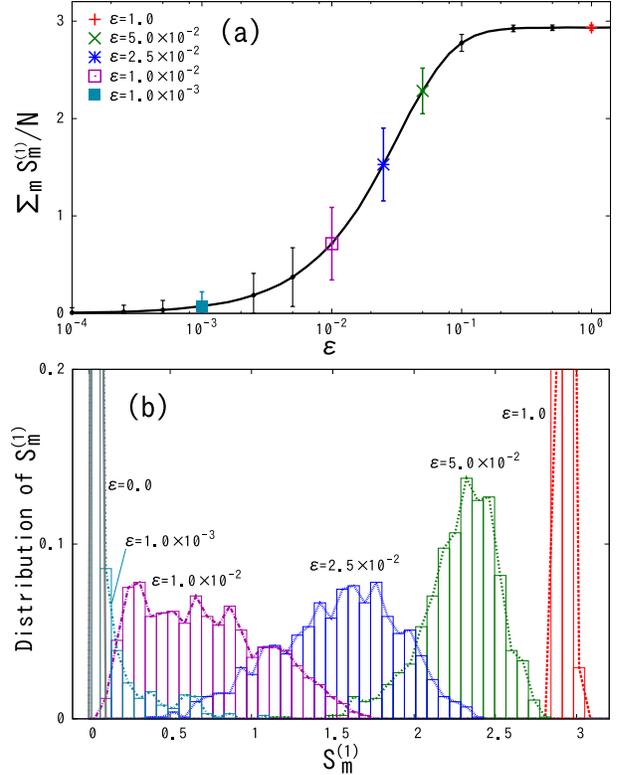}
\caption{
\label{fig2}(a) 
The mean EE of all eigenstates $\sum_{m=1}^N \Sm1/N$, 
and the standard deviation of EE of engenstates around the mean EE 
which is indicated by error bars, where $N_1=32$.
(b) The distribution function of EE of eigenstates exhibiting the anomalously large fluctuation
 as $\eps$ is increased from the non-entangled regime to the fully entangled regime.
}
\end{center}
\end{figure}

\section{Development of entanglement and lifetime}

In the CKR the coupling parameter $\eps$ 
sensitively 
controls its statistical properties. If the standard map with unbounded phase space is used for each KR,
the classical chaotic diffusion is recovered as $\eps$ is increased beyond a threshold proportional 
to $\hbar$. In a previous paper we showed that a similar transition occurs also for bounded finite 
dimensional CKR and the lifetime of 
correspondence with classical diffusive behavior 
is drastically enhanced \cite{matsuiletter}.

In Fig.\ref{fig1}(a) we show some examples of $\msd(\tau)$ as a function 
of $\tau$, which indicate the tendency that the time scale on which 
$\msd(\tau)$ follows the classical $\msd(\tau)$ increases with $\eps$. 
In Fig.\ref{fig1}(b) we depict how the average lifetime increases with 
the coupling strength $\eps$, demonstrating that the time scale of 
correspondence with classical diffusive behavior increases with $\eps$.

\begin{figure*}[htbp]
\begin{center}
\includegraphics[width=15.5cm] 
{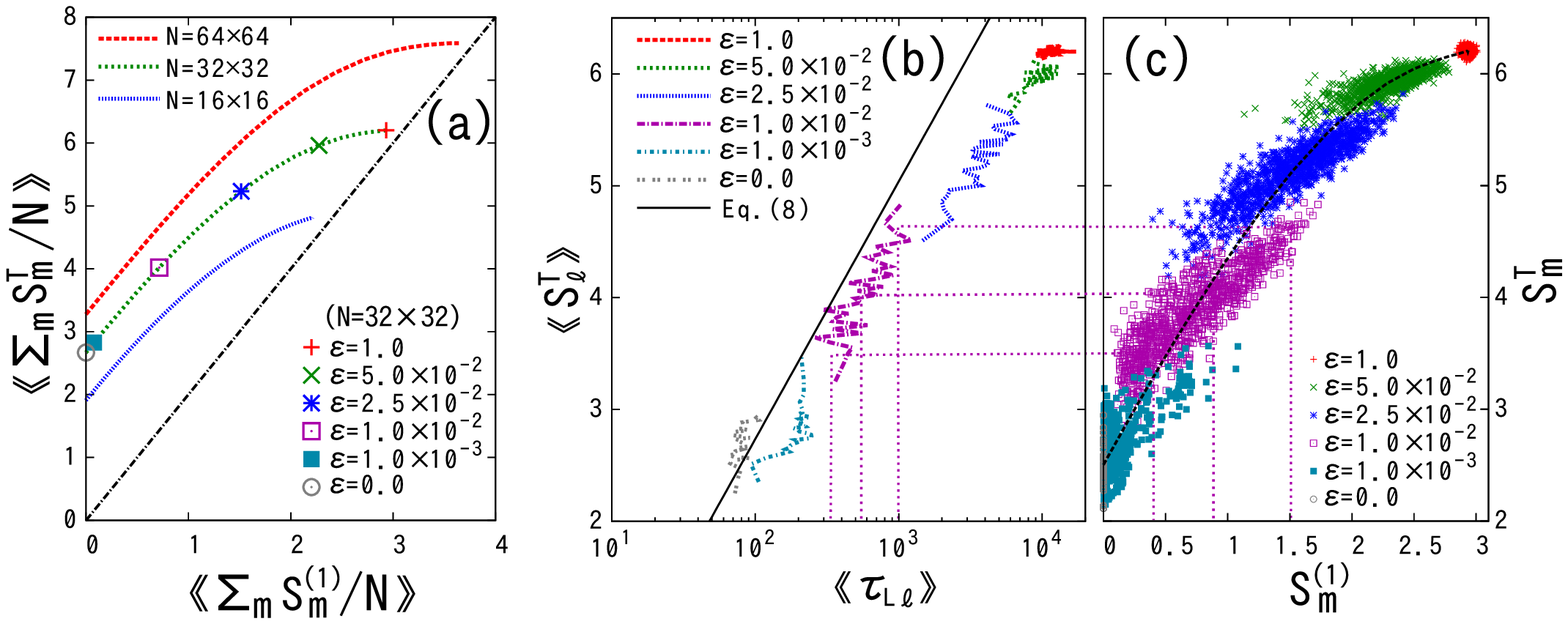}
\caption{
\label{fig3} (a) The correlation between the mean EE $\sum_{m=1}^N S^{(1)}_m /N$ 
and the mean TE $\sum_{m=1}^N S^T_m / N$, where $\ll~~\rr$ 
means values are ensemble-averaged. (b) Correlation between the average lifetime 
and the mean TE for various $\eps$. The straight line indicates $\tlife \propto e^{S^T}=B$. 
(c) Plots of EE vs TE i.e., $(\Sm1,S^T_m)$ of individual eigenfunction $|m\>$ for various $\eps$.
The black curve indicates correlation between the mean values $\sum_{m=1}^N S^T_m/N$ and 
$\sum_{m=1}^N \Sm1/N$ (the green curve in (a)). $\tlife$ in (b) of $\eps=0.01$(purple), 
for example, is correlated with EE of (c) through TE as is indicated by purple dotted lines. 
} 
\end{center}
\end{figure*}

We are interested in the relation between growth of the irreversibility and development 
of the entanglement in the CKR, especially for the bounded KR. The von-Neumann entropy is a 
standard tool to measure the degree of entanglement quantitively \cite{couplerotorentangle}.   
We use it to measure the 
entanglement between the two KR's. Considering the reduced density operator $\rhom1=\Tr_2|m\>\<m|$ 
traced over the KR2 the von-Neumann entropy for the reduced $\rhom1$, namely
\begin{eqnarray}
   \Sm1=-\Tr_1(\rhom1\log\rhom1)
\end{eqnarray}
is the entanglement entropy(EE) for KR1.  We show in Fig.\ref{fig1}(b) the mean EE, 
$\sum_{m=1}^N \Sm1/N$, as a function of $\eps$, which is averaged over the ensemble of 
the perturbation potential parameter $q_{iR}$. 

It is evident that the average lifetime is strongly correlated with the mean EE. 
We can not, however, know what kind of physical process is happening in the system from EE alone.
We explore more closely the connection between EE and $\tlife$.  
In the transition regime between the unentangled regime and the fully-entangled one, 
a quite interesting phenomenon is observed numerically: 
the distribution of EE for individual eigenstates 
spreads over the whole range of values from 0 to the fully entangled value. 
The mean EE and the standard deviation of EE around 
the mean value are shown in Fig.\ref{fig2}(a), and the distribution function of EE are depicted
 in Fig.\ref{fig2}(b). The spread of the EE distribution 
indicates a critical-like phenomenon in the transition regime,
which means that the degree of entanglement fluctuates anomalously from state to state 
in the range between the unentangled value and the fully-entangled value, 
although we could not confirm the presence of a critical phenomenon in the rigorous sense
(namely, critical slowing down and critical divergence of some physical quantities).
Then we can expect that such an anomalous fluctuation should be apparent in the 
fluctuation of the lifetime, and the more entangled eigenstate should have longer lifetime. 
However, the following question arises: 
for individual eigenstates, does the correlation between lifetime and EE exist? 
Unfortunately, the accidental fluctuation of lifetime is in general so large that it is very 
difficult to numerically extract the correlation between the EE and the lifetime 
for individual eigenstates.

\section{Correlation between entanglement and lifetime at the level of 
 individual eigenstates} 

To supplement the EE, we introduce a more intuitively comprehensible entropy-like quantity.
Since $\v(\p_1)$ is free from KR2, the number of eigenstates connected by $\v$ increases
as the entanglement between KR1 and KR2 is enhanced. 
We therefore define the {\it transition rate entropy} (TE) $S^T_m$ representing the variety of 
transitions from the state $|m\>$ due to the perturbation $\v$ as 
\begin{eqnarray}
  S^T_m=-\sum_{n}t_{mn}\log t_{mn}, 
\end{eqnarray}
where $t_{mn}=|\<m|\hat{v}|n\>|^2/\sum_{n'} |\<m|\hat{v}|n'\>|^2$ is the normalized transition rate.
Its mean value $\sum_{m=1}^N S^T_m/N$ as a function of $\eps$  
has a definite correlation with the mean EE $\sum_{m=1}^N \Sm1/N$ as is displayed in Fig.\ref{fig3}(a).
To elucidate the relation between 
the EE and the lifetime, we can use the TE introduced above as a mediating quantity between them. 
We can expect that TE is strongly correlated with $\tlife$ of individual eigenstate. This is based 
upon the following theoretical considerations. First, we note that the TE $S^T_m$ represents 
the number of eigenstates which are connected with  the eigenstate $|m\>$, 
by the relation $B_m= e^{S^T_m}$.
Next, in the previous paper we showed that if we could suppose that the eigenstate $|m\>$
are connected by $\v$ with all other eigenstates almost equally, then the  
lifetime of the superposed state $|\Psi_0\>=\sum_{m=1}^M C_m|m\>$ is given by 
\begin{eqnarray}
\label{tlife}
        \tlife \sim \frac{2D\corr(0)MN}{\pi^2D_{\omega}^{(cl)}},
\end{eqnarray}
which predicts a seemingly strange feature that the lifetime depends on
$M$ i.e., the number of eigenstates forming the initial state $|\Psi_0\>$, 
where $\corr(0)=\sum_{m=1}^M |C_m|^2 \sum_{n=1}^N | \<m|\v|n\>|^2$ 
\cite{matsuiletter}.
Equation (\ref{tlife}) explains the numerical results quite well in the fully 
entangled regime.  This result can be extended in the transition 
regime if we suppose that $\v$ is dominantly connected only with $B=\e^{S^T}$ 
eigenstates among the $N$ eigenstates, and moreover the $M$ states 
forming $|\Psi_0\>$ are so chosen as to have almost the same value of the
transition rate entropy $S^T$ (or $B$).
Then $N$ in Eq.(\ref{tlife}) can be replaced by $B_m=e^{S^T_m}$ and the lifetime 
of such a $|\Psi_0\>$ satisfies 
$\tlife \propto BM = \e^{S^T} M$. 
Thus we may expect that $S^T=\log B$ is straightforwardly connected with $\tlife$. 
In the short limit of the correlation time, 
the classical diffusion coefficient is approximated as $D_\omega^{(cl)} \sim D \corr(0)$, 
then the Eq.(\ref{tlife}) becames 
\begin{eqnarray}
\label{tlifeTE}
        \tlife \sim \frac{2M}{\pi^2} e^{S^T},
\end{eqnarray}
as is the case for the coupled Arnold's cat maps.
To confirm this conjecture numerically we executed the following procedures: 
(A) Sort $N=N_1^2$ eigenfunctions in order of the decreasing TE.
(B) The sorted eigenfunctions are grouped into $N/M$ sets in descending order 
which are numbered as $\ell=1,2,...N/M$, where each of the sets contain $M$ eigenfunctions.
(C) Superpose $M$ eigenfunctions with almost the same TE to form the representative state 
$|\Psi_0^{(\ell)}\>$ of $\ell$-th group and measure its ${\tlife}_\ell$ and compute the mean TE, say $S^T_\ell$.  
(D) Take the ensemble average of ${\tlife}_\ell$ and $S^T_\ell$ over the potential parameter 
$q_{iR}~(i=1,2)$.
We specifically take $M=N_1$.  We show in Fig.\ref{fig3}(b) the plots 
$(\<\!\<{\tlife}_\ell\>\!\>, ~\<\!\<S^T_\ell\>\!\>)$
for several $\eps$ in the transition regime. The plotted points are connected by lines 
to guide the eye. It is evident that the lifetime is almost proportional to $e^{S^T}$.

On the other hand, as shown in Fig.\ref{fig3}(a) the mean values of EE and TE are highly 
correlated. The question is whether or not the correlation exists even at the level of 
individual eigenstates. Figure \ref{fig3}(c) we examined the correlation between EE and TE 
for an individual 
eigenstate. 
It is evident that there exists a clear positive correlation 
between EE and TE in each of the anomalously fluctuating sets with different values of $\eps$.
Combining Fig.\ref{fig3}(b) and (c), we can claim that, through the 
transition rate entropy, the existence of a strong positive correlation between the 
entanglement entropy and the lifetime of irreversibility is evident at the level of 
individual eigenfunctions.

\section{Conclusion}
We showed that, by taking the classically chaotic coupled quantum kicked 
rotors as examples, the lifetime of correspondence with classical decay 
of correlation, which indicates the maximal time scale on which the quantum 
time-irreversibility 
is observable, can be measured even for individual eigenstates.  The 
lifetime is strongly correlated with the degree of entanglement between the kicked rotors.
The lifetime varies from eigenstate to eigenstate 
particularly in the threshold regime at the onset of full entanglement.
But by introducing the transition rate entropy, we succeed in showing that 
the anomalous fluctuation of 
entropy has a definite positive correlation with 
the fluctuation of lifetime for the eigenstates. 

The lifetime used here plays the role of an order parameter for investigating 
the cooperative nature of the critical phenomena in the transition process 
through which time-irreversible property is self-organized in {\it bounded} 
quantum systems, which has not been examined yet.

\acknowledgments
This work is partly supported by Japanese people's tax via JPSJ KAKENHI 15H03701,
and the authors would like to acknowledge them.
They are also very grateful to Kankikai, Dr.S.Tsuji, and  Koike memorial
house for using the facilities during this study.

\end{document}